# Expansion formulae for one- and two-center charge densities over complete orthonormal sets of exponential type orbitals and their use in evaluation of multicenter-multielectron integrals

I.I.Guseinov

*Department of Physics, Faculty of Arts and Sciences, Onsekiz Mart University, Çanakkale, Turkey*

**Abstract**

The series expansion formulae are established for the one- and two-center charge densities over complete orthonormal sets of $\Psi^{\alpha}$-exponential type orbitals ($\alpha = 1, 0, -1, -2, ...$) introduced by the author. Three-center overlap integrals of $\Psi^{\alpha}$ appearing in these relations are expressed through the two-center overlap integrals between $\Psi^{\alpha}$-orbitals. The general formulae obtained for the charge densities are utilized for the evaluation of arbitrary multicenter-multielectron integrals occurring when the complete orthonormal sets of $\Psi^{\alpha}$-exponential type orbitals are used as basis functions in the Hartree-Fock-Roothaan and explicitly correlated methods. The relationships for charge densities and multicenter-multielectron integrals obtained are valid for the arbitrary quantum numbers, screening constants and location of $\Psi^{\alpha}$-orbitals.

## 1. Introduction

In electronic structure calculations of molecules, the Slater type orbitals (STO) and complete orthonormal sets of $\Psi^{\alpha}$-exponential type orbitals ($\Psi^{\alpha}$-ETO) are viable alternative to Gaussian type orbitals (GTO). The GTO should be replaced by STO or $\Psi^{\alpha}$-ETO because they are more physical basis functions. The STO and $\Psi^{\alpha}$-ETO are able to satisfy the cusp condition at the nuclei [1] and decrease exponentially at the large distances [2], therefore, they behave as exact eigenstates of Hamiltonians of atomic and molecular systems. However, the STO functions are not orthogonal with respect to the principal quantum numbers that creates some difficulties in molecular electronic structure calculations. Thus, the necessity for using the complete orthonormal sets of $\Psi^{\alpha}$-ETO as basis functions arises [3]. A large number of different sets of molecular orbitals can be obtained for the different values of $\alpha$ ($\alpha = 1, 0, -1, -2, ...$), which have the further advantage that the basis sets of $\Psi^{\alpha}$-ETO, which are required, can be chosen properly according to the nature of the problems under consideration. This is rather important because the choice of the basis sets will determine the rate of convergence of series expansions arising in molecular electronic structure calculations.

Therefore, the complete orthonormal sets of $\Psi^{\alpha}$-ETO with different values of indices $\alpha$ could be universally used in electronic structure calculations of atoms and molecules.

The aim of this paper is to derive the general expansion formlae for the two-center $\Psi^{\alpha}$-ETO charge densities in terms of $\Psi^{\alpha}$-ETO at a third center and to evaluate the multicenter-multielectron integrals over $\Psi^{\alpha}$-ETO. We notice that the method used in this work presents the development of previous paper [5] in which the formulae for expansion and multicenter-multielectron integrals over integer and noninteger n-STO have been established.

**2. Definitions and basic formulas**

The $\Psi^{\alpha}$-ETO are determined as [3]

$$\Psi^{\alpha}_{nlm}(\zeta,\vec{r}) = (-1)^{\alpha}\left[\frac{(2\zeta)^{3}(q-p)!}{(2n)^{\alpha}(q!)^{3}}\right]^{1/2} x^{l} e^{-x/2} L^{p}_{q}(x) S_{lm}(\theta,\varphi), \tag{1}$$

where $\alpha = 1,0,-1,-2,...$, $p = 2l+2-\alpha$, $q = n+l+1-\alpha$ and $x = 2\zeta r$; $L^{p}_{q}(x)$ and $S_{lm}(\theta,\varphi)$ are the generalized Laguerre polynomials and the complex ($S_{lm} \equiv Y_{lm}$) or real spherical harmonics, respectively. We notice that the definition of phases in this work for the complex spherical harmonics ($Y^{*}_{lm} = Y_{l-m}$) differs from the Condon-Shortley phases [4] by sign factor $(-1)^{m}$.

The $\Psi^{\alpha}$-ETO are orthonormal with respect to the weight function $(n/\zeta r)^{\alpha}$,

$$\int \Psi^{\alpha\,*}_{nlm}(\zeta,\vec{r})\overline{\Psi}^{\alpha}_{n'l'm'}(\zeta,\vec{r})d^{3}\vec{r} = \delta_{nn'}\delta_{ll'}\delta_{mm'}, \tag{2}$$

where

$$\overline{\Psi}^{\alpha}_{nlm}(\zeta,\vec{r}) = (n/\zeta r)^{\alpha}\,\Psi^{\alpha}_{nlm}(\zeta,\vec{r}). \tag{3}$$

The purpose of the present study is to derive the analytical relations for $\Psi^{\alpha}$-charge densities and multicenter-multielectron integrals defined by

$$\rho^{\alpha}_{pp'}(\zeta,\vec{r}_{b};\zeta',\vec{r}_{c}) = \Psi^{\alpha}_{p}(\zeta,\vec{r}_{b})\Psi^{\alpha\,*}_{p'}(\zeta',\vec{r}_{c}) \tag{4}$$

$$\begin{aligned} &I^{\alpha}_{p_1p'_1,p_2p'_2,...,p_sp'_s}\left(\zeta_1,\zeta'_1,0,\vec{R}_{ac};\zeta_2,\zeta'_2,\vec{R}_{ab},\vec{R}_{ad};...;\zeta_s,\zeta'_s,\vec{R}_{ae},\vec{R}_{af}\right)\\ &= \int \hat{F}^{(s)}\rho^{\alpha\,*}_{p_1p'_1}(\zeta_1,\vec{r}_{a1};\zeta'_1,\vec{r}_{c1})\,\rho^{\alpha}_{p_2p'_2}(\zeta_2,\vec{r}_{b2};\zeta'_2,\vec{r}_{d2})...\rho^{\alpha}_{p_sp'_s}(\zeta_s,\vec{r}_{es};\zeta'_s,\vec{r}_{fs})dv_1dv_2...dv_s, \end{aligned} \tag{5}$$

where $p \equiv nlm$, $p' \equiv n'l'm'$, $p_i \equiv n_il_im_i$, $p'_i \equiv n'_il'_im'_i$, $s = 1,2,3,...$ and $\hat{F}^{(s)}$ is the arbitrary s-electron operator (see, e.g., Refs. [6-8]). The operator $\hat{F}^{(s)}$ might, in general, depend on coordinates of electrons and nuclei, and their derivatives. For example, in the case of Hartree-Fock-Roothaan equations for molecules, the operators $\hat{F}^{(1)}$ and $\hat{F}^{(2)}$ are determined by

$$\hat{F}^{(1)} = -\frac{1}{2}\nabla_1^2 - \frac{Z_g}{r_{g1}} \tag{6}$$

$$\hat{F}^{(2)} = \frac{1}{r_{21}}, \tag{7}$$

where $g = a, c, b, d, ..., e, f$. For nuclear attraction and electron repulsion there are one-, two-, three- and one-, two-, three-, four-center integrals, respectively. Thus, in general, the number of centers involved in Eq. (5) might be higher than 2s. The electronic coordinates can be separated from the nuclear coordinates using the unsymmetrical and symmetrical one-range addition theorems presented in previous paper [9]. The applicability of these theorems to the study of electronic structure of molecules has been demonstrated in Refs. [10, 11].

### 3. Expansions of $\Psi^{\alpha}$ -charge densities

In order to evaluate the multicenter- multielectron integrals with $\Psi^{\alpha}$ -ETO, we first derive the general expansion formula for the two-center $\Psi^{\alpha}$ -charge densities in terms of $\Psi^{\alpha}$ -ETO at a third center. Taking into account Eqs. (1)-(3) and the properties of $\Psi^{\alpha}$ -ETO we obtain:

$$\rho_{pp'}^{\alpha}\left(\zeta, \vec{r}_b; \zeta', \vec{r}_c\right) = \frac{1}{\sqrt{4\pi}} \sum_{\mu=1}^{\infty} \sum_{\nu=0}^{\mu-1} \sum_{\sigma=-\nu}^{\nu} S_{pp'q}^{\alpha}{}^{*}\left(\zeta, \zeta', z; \vec{R}_{ab}, \vec{R}_{ac}\right) \Psi_q^{\alpha}\left(z, \vec{r}_a\right). \tag{8}$$

Here, $S_{pp'q}^{\alpha}\left(\zeta, \zeta', z; \vec{R}_{ab}, \vec{R}_{ac}\right)$ are the three-center overlap integrals of $\Psi^{\alpha}$ -ETO defined by

$$\begin{aligned} S_{pp'q}^{\alpha}\left(\zeta\zeta' z; \vec{R}_{ab}, \vec{R}_{ac}\right) &= \sqrt{4\pi} \int \bar{\Psi}_q^{\alpha}\left(z, \vec{r}_a\right) \Psi_p^{\alpha*}\left(\zeta, \vec{r}_b\right) \Psi_{p'}^{\alpha}\left(\zeta', \vec{r}_c\right) d^3 \vec{r} \\ &= \sqrt{4\pi} \int \bar{\Psi}_q^{\alpha}\left(z, \vec{r}\right) \Psi_p^{\alpha*}\left(\zeta, \vec{r} - \vec{R}_{ab}\right) \Psi_{p'}^{\alpha}\left(\zeta', \vec{r} - \vec{R}_{ac}\right) d^3 \vec{r}, \end{aligned} \tag{9}$$

where $\vec{r} = \vec{r}_a$, $\vec{r} - \vec{R}_{ab} = \vec{r}_b$, $\vec{r} - \vec{R}_{ac} = \vec{r}_c$, $q \equiv \mu\nu\sigma$ and $z = \zeta + \zeta'$. Thus, the $\Psi^{\alpha}$ -charge densities are expressed through the multicenter overlap integrals of three $\Psi^{\alpha}$ -ETO.

For the evaluation of multicenter overlap integrals of three $\Psi^{\alpha}$ -ETO we use the following expansion and one-range addition theorems [12]:

$$\bar{\Psi}_p^{\alpha*}\left(\zeta, \vec{r}\right) \Psi_{p'}^{\alpha}\left(\zeta', \vec{r}\right) = \frac{(2z)^{3/2}}{\sqrt{4\pi}} \sum_{\mu=1}^{n+n'-1} \sum_{\nu=0}^{\mu-1} \sum_{\sigma=-\nu}^{\nu} \bar{B}_{pp'}^{\alpha q}(\eta) \bar{\Psi}_q^{\alpha*}\left(z, \vec{r}\right) \tag{10}$$

$$\Psi_p^{\alpha}\left(\zeta, \vec{r} - \vec{R}_{ab}\right) = \sum_{\mu=1}^{\infty} \sum_{\nu=0}^{\mu-1} \sum_{\sigma=-\nu}^{\nu} \bar{S}_{qp}^{\alpha}\left(\vec{G}_{ab}\right) \Psi_q^{\alpha}\left(\zeta, \vec{r}\right), \tag{11}$$

where $\eta = \zeta/\zeta'$, $\vec{G}_{ab} = 2\zeta \vec{R}_{ab}$ and

$$\bar{B}_{pp'}^{\alpha q}(\eta) = \frac{\sqrt{4\pi}}{(2z)^{3/2}} \int \bar{\Psi}_p^{\alpha*}\left(\zeta, \vec{r}\right) \Psi_{p'}^{\alpha}\left(\zeta', \vec{r}\right) \Psi_q^{\alpha}\left(z, \vec{r}\right) d^3 \vec{r} \tag{12}$$

$$\bar{S}_{qp}^{\alpha}\left(\vec{G}_{ab}\right) \equiv \bar{S}_{qp}^{\alpha}\left(\zeta,\zeta;\vec{R}_{ab}\right) = \int \bar{\Psi}_{q}^{\alpha *}\left(\zeta,\vec{r}\right)\Psi_{p}^{\alpha}\left(\zeta,\vec{r}-\vec{R}_{ab}\right)d^{3}\vec{r}\ . \tag{13}$$

The analytical relations for linear combination coefficients $\bar{B}_{pp'}^{\alpha q}(\eta)$ and two-center overlap integrals $\bar{S}_{qp}^{\alpha}\left(\vec{G}_{ab}\right)$ are presented in [12].

Now we take into account Eqs. (10) and (11) in (9). Then, we obtain for overlap integrals of three $\Psi^{\alpha}$-ETO the following relations through the overlap integrals with two $\Psi^{\alpha}$-ETO:

for three-center cases $\left(a \neq b, a \neq c, b \neq c\right)$

$$S_{pp'q}^{\alpha}\left(\zeta,\zeta',z;\vec{R}_{ab},\vec{R}_{ac}\right) = \left(2z_{1}\right)^{\frac{3}{2}} \sum_{\mu_1=1}^{\infty}\sum_{\nu_1=0}^{\mu_1-1}\sum_{\sigma_1=-\nu_1}^{\nu_1} \bar{S}_{q_1 p}^{\alpha\,*}\left(\vec{G}_{ab}\right) \sum_{\mu_2=1}^{\mu+\mu_1-1}\sum_{\nu_2=0}^{\mu_2-1}\sum_{\sigma_2=-\nu_2}^{\nu_2} \bar{B}_{qq_1}^{\alpha q_2}\left(\eta_1\right)\bar{S}_{q_2 p'}^{\alpha}\left(z_1,\zeta';\vec{R}_{ac}\right) \tag{14a}$$

$$S_{pp'q}^{\alpha}\left(\zeta,\zeta',z;\vec{R}_{ab},\vec{R}_{ac}\right) = \left(2z_{2}\right)^{\frac{3}{2}} \sum_{\mu_1=1}^{\infty}\sum_{\nu_1=0}^{\mu_1-1}\sum_{\sigma_1=-\nu_1}^{\nu_1} \bar{S}_{q_1 p'}^{\alpha}\left(\vec{G}'_{ac}\right) \sum_{\mu_2=1}^{\mu+\mu_1-1}\sum_{\nu_2=0}^{\mu_2-1}\sum_{\sigma_2=-\nu_2}^{\nu_2} \bar{B}_{qq_1}^{\alpha q_2}\left(\eta_2\right)\bar{S}_{q_2 p}^{\alpha\,*}\left(z_2,\zeta;\vec{R}_{ab}\right), \tag{14b}$$

for two-center cases $\left(a \neq b,\ a \neq c,\ b \equiv c;\ a \equiv b,\ a \neq c;\ a \neq b,\ a \equiv c\right)$

$$S_{pp'q}^{\alpha}\left(\zeta,\zeta',z;\vec{R}_{ab},\vec{R}_{ac}\right) = \sum_{\mu_1=1}^{n+n'-1}\sum_{\nu_1=0}^{\mu_1-1}\sum_{\sigma_1=-\nu_1}^{\nu_1} S_{pp'q_1}^{\alpha}\left(\zeta\zeta' z\right)\bar{S}_{qq_1}^{\alpha\,*}\left(z,z;\vec{R}_{ab}\right) \tag{15a}$$

$$S_{pp'q}^{\alpha}\left(\zeta,\zeta',z;0,\vec{R}_{ac}\right) = \left(2z_{1}\right)^{\frac{3}{2}} \sum_{\mu_1=1}^{\mu+n-1}\sum_{\nu_1=0}^{\mu_1-1}\sum_{\sigma_1=-\nu_1}^{\nu_1} \bar{B}_{qp}^{\alpha q_1}\left(\eta_1\right)\bar{S}_{q_1 p'}^{\alpha}\left(z_1,\zeta';\vec{R}_{ac}\right) \tag{15b}$$

$$S_{pp'q}^{\alpha}\left(\zeta,\zeta',z;\vec{R}_{ab},0\right) = \left(2z_{2}\right)^{\frac{3}{2}} \sum_{\mu_1=1}^{\mu+n'-1}\sum_{\nu_1=0}^{\mu_1-1}\sum_{\sigma_1=-\nu_1}^{\nu_1} \bar{B}_{qp'}^{\alpha q_1}\left(\eta_2\right)\bar{S}_{q_1 p}^{\alpha\,*}\left(z_2,\zeta';\vec{R}_{ab}\right), \tag{15c}$$

for one-center case $\left(a \equiv b \equiv c\right)$

$$S_{pp'q}^{\alpha}\left(\zeta,\zeta',z;0,0\right) = \begin{cases} S_{pp'q}^{\alpha}\left(\zeta,\zeta',z\right) & for\ 1 \le \mu \le n+n'-1 \quad (16a) \\ 0 & for\ \ \mu > n+n'-1 \quad (16b) \end{cases}$$

$$S_{pp'q}^{\alpha}\left(\zeta,\zeta',z\right) = \left(2z\right)^{\frac{3}{2}} \bar{B}_{pp'}^{\alpha q}\left(\eta\right), \tag{16c}$$

where $\vec{G}'_{ac} = 2\zeta'\vec{R}_{ac}$, $z_1 = z + \zeta$, $\eta_1 = \frac{z}{\zeta}$, $z_2 = z + \zeta'$, $\eta_2 = \frac{z}{\zeta'}$ and

$$\bar{S}_{pp'}^{\alpha}\left(\zeta,\zeta';\vec{R}_{ab}\right) = \int \bar{\Psi}_{p}^{\alpha *}\left(\zeta,\vec{r}\right)\Psi_{p'}^{\alpha}\left(\zeta',\vec{r}-\vec{R}_{ab}\right)d^{3}\vec{r}. \tag{17}$$

### 4. Expression for multicenter-multielectron integrals of $\Psi^{\alpha}$-ETO

According to the Hartree-Fock theory the matrix elements of s-electron operator (s=1,2,3,...) between the determinantal wave functions of molecules are expressed through the multicenter multielectron integrals defined by Eq. (5). For the evaluation of these integrals we use Eq. (6) for all the $\Psi^{\alpha}$-charge densities which occur in Eq. (5). Then, we obtain the following s-fold infinite expansion relation in terms of the overlap integrals and the s-electron basic integrals of $\Psi^{\alpha}$ :

$$\begin{aligned} I^{\alpha}_{p_1p_1',p_2p_2',...,p_sp_s'}(\zeta_1,\zeta_1',0,\vec{R}_{ac};\zeta_2,\zeta_2',\vec{R}_{ab},\vec{R}_{ad};...;\zeta_s,\zeta_s',\vec{R}_{ae},\vec{R}_{af}) &= \sum_{q_1q_2...q_s} J^{\alpha}_{q_1q_2...q_s}(z_1,z_2,...,z_s) \\ \times S^{\alpha *}_{p_1p_1'q_1}\left(\zeta_1,\zeta_1',z_1;0,\vec{R}_{ac}\right) S^{\alpha}_{p_2p_2'q_2}\left(\zeta_2,\zeta_2',z_2;\vec{R}_{ab},\vec{R}_{ad}\right) &... S^{\alpha}_{p_sp_s'q_s}\left(\zeta_s,\zeta_s',z_s;\vec{R}_{ae},\vec{R}_{af}\right), \end{aligned} \tag{18}$$

where $1\le\mu_i\le\infty,\ 0\le\nu_i\le\mu_i-1,\ -\nu_i\le\sigma_i\le\nu_i\quad(1\le i\le s)$ and the s-electron basic integrals of $\Psi^{\alpha}$-ETO are defined as

$$J^{\alpha}_{q_1q_2...q_s}(z_1,z_2,...,z_s)=\frac{1}{(4\pi)^{\frac{s}{2}}}\int \hat{F}^{(s)}\Psi^{\alpha *}_{q_1}\left(z_1,\vec{r}_1\right)\Psi^{\alpha}_{q_2}\left(z_2,\vec{r}_2\right)...\Psi^{\alpha}_{q_s}\left(z_s,\vec{r}_s\right)dv_1dv_2...dv_s. \tag{19}$$

The convergence of series expansions occurring in Eq. (18) is quaranteed because they have been derived with the help of complete orthonormal sets of $\Psi^{\alpha}$-ETO.

By the use of relation [3]

$$\Psi^{\alpha}_{nlm}\left(\zeta,\vec{r}\right)=\sum_{\mu=l+1}^{n}\omega^{\alpha l}_{n\mu}\chi_{\mu lm}(\zeta,\vec{r}), \tag{20}$$

where $\chi_{\mu lm}(\zeta,\vec{r})$ are the STO, Eq. (19) can be expressed through the s-electron basic integrals of STO (see Ref. [5]). Thus, the formulas obtained in our previous papers for the multicenter integrals of STO can be also used to calculate multicenter-multielectron integrals over complete orthonormal sets of $\Psi^{\alpha}$-ETO.

Taking into account the series expansion formulas obtained in this study, one can calculate all the multielectron molecular integrals arising in the determination of various properties for a given molecule when the complete orthonormal sets of $\Psi^{\alpha}$-ETO are used in Hartree-Fock-Roothaan and explicitly correlated methods. We notice that the two- and three-center overlap integrals with $\Psi^{\alpha}$-ETO appearing in Eq. (18) for the multicenter-multielectron integrals can be calculated using two-center ovelap integrals of STO for the computation of which efficient computer programs are available in our group (see Ref. [13] and references quoted therein to our papers for overlap integrals of STO).